\documentstyle[aps,amssymb]{revtex}

\begin{document}
\author{Mario Alberto Castagnino.}
\address{I.A.F.E. (Univ. de Bs. As.)}
\author{Adolfo Ram\'{o}n Ord\'{o}\~{n}ez.}
\address{I.F.I.R. / Fac. de Ciencias Exactas, Ing. y Agrim. (Univ. Nac. de Rosario)}
\author{Daniela Beatriz Emmanuele.}
\address{Fac. Cs. Exactas, Ing.y Agrim.U.N.R.}
\title{A general mathematical structure for the time-reversal operator.}
\date{December 27th., 2000}
\maketitle

\begin{abstract}
The aim of this work is the mathematical analysis of the physical
time-reversal operator and its definition as a geometrical structure{\bf , }%
in such a way that it could be generalized to the purely mathematical realm.
Rigorously, only having such a ``time-reversal structure'' it can be decided
whether a dynamical system is time-symmetric or not.{\it \ }The
``time-reversal structures'' of several important physical and mathematical
examples are presented, showing that there are some mathematical categories
whose objects are the (classical or abstract) ``time-reversal systems'' and
whose morphisms generalize the Wigner transformation.
\end{abstract}

\section{Introduction.}

The dynamics and the thermodynamics of both, classical and quantum physical
systems, are modelized by the mathematical theory of classical and abstract
dynamical systems. It is obvious that the {\it physical} notion of
``time-symmetric (or asymmetric) dynamical systems'' requires the definition
of a ``time-reversal operator'', $K$ \cite{7}$.$ In fact, every known model
of a physical dynamical system {\it has} some $K$ operator. E.g., the
dynamic of classical physical systems is described in the cotangent fiber
bundle $T^{*}(N)$ of its configuration manifold $N$, and therefore the
action of $K:T^{*}(N)\rightarrow T^{*}(N)$ is defined as

\begin{equation}
p_{q}\mapsto -p_{q}  \label{0.1}
\end{equation}
for any linear functional $p$ on $q\in N,$ or in coordinate's language: 
\begin{equation}
(q^{i},p_{i})\mapsto (q^{i},-p_{i})  \label{0.2}
\end{equation}
in a particular $(q^{i},p_{i})$ coordinate system.

In Quantum Mechanics there is the well known Wigner antiunitary operator $K$
defined through the complex conjugation in the position (wave functions)
representation: 
\begin{equation}
\psi (x,t)\mapsto \psi (x,-t)^{*}  \label{0.3}
\end{equation}

In the last few years it was demonstrated that ordinary Quantum Mechanics
(with no superselection sectors) can be naturally included in the
Hamiltonian formalism of its real K\"{a}hlerian differentiable manifold of
quantum states \cite{Cire0} \cite{Cire1} \cite{Cire2}. The latter one is the
real but infinite dimensional manifold of the associated projective space $%
{\bf P}({\cal H})$ of its Hilbert states space ${\cal H}$ \footnote{%
We should remember the fact that ordinary pure quantum states are not {\it %
vectors} $\psi $ (normalized or not) of a Hilbert space ${\cal H}$, but rays
or {\it projective equivalence classes of vectors} $[\psi ]\in {\bf P}({\cal %
H})$.}. From this point of view, $K:{\bf P}({\cal H})\rightarrow {\bf P}(%
{\cal H})$ acts as the cannonical projection to the quotient of (\ref{0.3}) 
\begin{equation}
\lbrack \psi (x,t)]\mapsto \left[ \psi (x,-t)^{*}\right]  \label{0.4}
\end{equation}

Moreover, this result has been generalized to more general quantum systems
through its characteristic C*-algebra $A$, and its pure quantum states space 
$\partial K(A)$ turns out to be a projective K\"{a}hler bundle over the
spectrum $\widehat{A},$ whose fiber over the class of a state $\psi ,$ is
isomorphic to ${\bf P}({\cal H}_{\psi }),$ being ${\cal H}_{\psi }$ its GNS
(Gelfand-Naimark-Segall) representation's space \cite{Cire3}. More recently
these authors have relaxed the K\"{a}hlerian structure, showing ''minimal''
mathematical structures involved in the quantum principles of superposition
and uncertainty, with the aim of considering non linear extensions of
quantum mechanics \cite{Cire4}.

But, what happens in more general dynamical systems? Some of them -such as
the Bernouilli systems and certain Kolmogorov-systems \cite{2}- are purely
mathematical. Nevertheless, the notion of time-symmetry seems to make sense
also for them. So, it would be interesting to know what kind of mathematical
structures are involved in these systems.

Our aim is to show that:

1.-{\it The mere existence of the time reversal operator is a mathematical
structure }consisting of a non trivial involution $K$ of the states space of
a general system (with holonomic constraints) $M,$ which splits into a $K$%
-invariant submanifold (coordinatized by ${\it q}^{i}$) and whose
complementary set (the field of the effective action of $K,$ coordinatized
by ${\it p}_{i}$) is a manifold with the same dimension of $M.$ This
structure is logically independent of the symplectic one \cite{1}, which
doesn't require such a splitting at all. In fact, the essence of symplectic
geometry, as its own etymology shows, is the ''common enveloping'' of $q$
and $p$, loosing any ''privilege'' between them. Actually, {\it this }$K$%
{\it -structure is defined by the action of that part of the complete
Galilei group -including the time-reversal- which is allowed to act on the
phase space manifold }$M$ {\it by the constraints. }In fact, only on the
homogeneous Euclidean phase space $M^{\prime }={\Bbb R}^{6n},$ it is
possible to have the transitive action \cite{4} of the complete Galilei group%
{\it . }

2.-{\it It is possible to define generalized and purely mathematical
``time-reversals'' }allowing a generalization of the notion of
``time-symmetry'' for a wider class of dynamical systems, including all
Bernouilli systems. In fact there are mathematical categories whose objects
are the time-reversal (classical or abstract) systems $(M,K)$ and whose
morphisms generalize the Wigner transformation \cite{Wigner}{\it . }

3.-When the states space has additional structures, {\it there is a
possibility of getting a richer time-reversal compatible with these
structures. }For example, in Classical Mechanics the canonical $K$ is a
symplectomorphism of phase space, and the Wigner quantum operator is
compatible with the K\"{a}hlerian structure of{\it \ }${\bf P}({\cal H}).$
At first sight (\ref{0.2}) is quite similar to (\ref{0.3}) and it seems to
be some kind of ''complex conjugation'' (and the even dimensionality of
phase space reinforces this idea). We will prove this fact, namely, the
existence of an almost complex structure $J$ with respect to which $K$ is an
almost complex time-reversal. This increases the analogy with Quantum
Mechanics, where the strong version of the Heisemberg Uncertainty Principle, 
\cite{Cire1} \cite{Cire2} is equivalent to the existence of a complex
structure $J,$ by means of which the Wigner time-reversal is defined.

4.-{\it It is possible to make a definition of time-reversal systems so
general }that it includes among its examples the real line, the Minkowski
space-time, the cotangent fiber bundles, the quantum systems, the classical
densities function space, the quantum densities operator space, the
Bernouilli systems, etc.

The paper is organized as follows: In section 2, the general theory of{\it \ 
}{\bf reversals }and{\bf \ time-reversal systems} is developed. In section
3, the theory of {\bf abstract} {\bf reversals }and {\bf abstract\
time-reversal systems} is given. In section 4, many important examples of
time-reversal systems are given{\it .} In section 5, we give the abstract
time-reversal of Bernouilli systems and we show explicitly the geometrical
meaning of our definitions for the Baker's transformation.

\section{Reversals and time-reversal systems}

{\bf Definition:} Let $M$ be a real paracompact, connected, finite or
infinite-dimensional differentiable manifold, and let $K:M\rightarrow M$ be
a diffeomorphism. We will say that $K$ is a {\bf reversal} on $M,$ and that $%
(M,K)$ is a {\bf reversal system} if the following conditions are satisfied:

(r.1) $K$ is an involution, i.e. $K^2=I_M$

(r.2) The set $N$ of all fixed points of $K$ is an immersed submanifold of $%
M,$ such that $M-N$ is a connected or disconnected submanifold of the same
dimension of $M$. (In particular, this implies that $M$ is a non trivial
involution, i.e. $K\neq I_{M},$ the identity function on $M$)

We will say that $N$ is the {\bf invariant} {\bf submanifold} of the
reversal system.

{\bf Definition:} Let $(M,K)$ be a reversal system. We will say that $M$ is $%
K${\bf -orientable} if $M-N$ is composed of two diffeomorphic connected
components $M_{+}$ y $M_{-}$, and if 
\begin{equation}
K(M_{+})=M_{-}\;,\;\text{and }K(M_{-})=M_{+}  \label{1.0}
\end{equation}

$M$ is $K${\bf -oriented} when -conventionally or arbitrarily- one of these
components is selected as {\bf ``positively oriented''}. In that case, $K$
changes the $K${\bf -}orientation of $M.$

If there is a complex (or almost complex) structure $J$ on $M$ (and
therefore $J^{\prime }=-J$ is another one) and if, in addition, $K$
satisfies:

(c.r.) $K$ is complex (or almost complex) as a map from $(M,J)$ to $%
(M,J^{\prime })=(M,-J)$, i.e. : 
\begin{equation}
K_{*}\circ J=-J\circ K_{*}  \label{1.1}
\end{equation}
\noindent we will say that $K$ is a {\bf complex (or almost complex)
reversal, or a conjugation. }

Similarly, if a symplectic (or almost symplectic) 2-form $\omega $ is given
on $M$ \footnote{%
In the infinite-dimensional case we require $\omega $ to be {\it strongly
non-degenerate }\cite{Cire4} in the sense that the map $X\mapsto \omega
(X,.) $ is a toplinear isomorphism.} (and therefore $\omega ^{\prime
}=-\omega $ is another one) and if, in addition to (r.1) y (r.2), $K$
satisfies:

(s.r.) $K$ is a symplectomorphism from $(M,\omega )$ to $(M,\omega ^{\prime
})=(M,-\omega )$, i.e. : 
\begin{equation}
K^{*}\omega =-\omega  \label{1.2}
\end{equation}

\noindent we will say that $K$ is a {\bf symplectic (or almost symplectic)
reversal. }If we have a symplectic (or almost symplectic) reversal system $%
(M,\omega ,K),$ then for every 
\[
m\in M:K_{*}:T_{m}(M)\rightarrow T_{m}(M) 
\]
is a (toplinear) isomorphism, and if 
\[
i:N\rightarrow M\text{ is the immersion}:i(q)=m 
\]
and $i_{*}(X_{q})=X_{i(q)}$ is the induced isomorphism, we can define an
almost complex structure $J$: 
\begin{eqnarray}
J\left( X_{i(q)}\right) &:&=Y_{m}\Leftrightarrow \omega \left(
X_{i(q)},Y_{m}\right) =1  \nonumber \\
J\left( Y_{m}\right) &:&=-X_{i(q)}  \label{1.2a}
\end{eqnarray}
that is to say, by defining the pairs of ``conjugate'' vectors (and
extending by linearity). Then 
\[
K_{*}\left( X_{i(q)}\right) =X_{i(q)}\;,\;K_{*}\left( Y_{m}\right) =-Y_{m} 
\]

When $(M,\omega ,J,g)$ is a K\"{a}hler (or almost K\"{a}hler) manifold, and $%
K$ satisfies the properties (r.1), (r.2), (c.r.) and (s.r.), we will say
that $K$ is a {\bf K\"{a}hlerian (or almost K\"{a}hlerian) reversal. }In
that case, $K$ is also an isometry 
\begin{equation}
K^{*}g=g  \label{1.3}
\end{equation}
with respect to the K\"{a}hler metric $g$ defined by:

\begin{equation}
g(X,Y)=-\omega (X,JY)\text{ for all vector fields }X\text{ and }Y.
\label{1.4}
\end{equation}

{\bf Definition:} Let $(M,K)$ be a reversal system such that there is a
class ${\cal F}$ of flows $\left( S_t\right) _{t\in {\Bbb R}}$ or / and
cascades $\left( S_t=S^t\right) _{t\in {\Bbb Z}}$ on $M$ such that, for any $%
m\in M,$ and any $t$ in ${\Bbb R}$ (or in ${\Bbb Z}$) satisfies: 
\begin{equation}
(K\circ S_t\circ K)(m)=S_{-t}(m)  \label{1.5}
\end{equation}

Then we will say that $K$ is a {\bf time-reversal }for ${\cal F}${\bf \ }on $%
M.$ (In Physics we can take ${\cal F}$ as a class of physical interest. For
example, in Classical Mechanics we can take the class of all Hamiltonian
flows over a fixed phase space and in Quantum Mechanics the class of
solutions of the Schr\"{o}dinger equation in a fixed states space, etc.)

Only having a time-reversal on $M,$ {\bf time-symmetric} {\bf (or asymmetric)%
} dynamical systems $\left( S_{t}\right) $ (flows $\left( S_{t}\right)
_{t\in {\Bbb R}}$ ; or cascades $\left( S_{t}=S^{t}\right) _{t\in {\Bbb Z}}$
) can be defined. In fact, $\left( S_{t}\right) $ will be considered as{\bf %
\ symmetric with respect to the time-reversal }$K${\bf ,} if it fulfills $%
\forall m\in M$ the condition (\ref{1.5}) (or {\bf asymmetric }if it
doesn't).

When $M$ is orientable (oriented) with respect to a time-reversal $K$, we
will say that it is {\bf time-orientable (oriented). }

{\bf Definition: }By a {\bf morphism} of the reversal system $(M,K)$ into $%
(M^{\prime },K^{\prime }),$ we mean a differentiable map $f$ of $M$ into $%
M^{\prime }$ such that 
\begin{equation}
f\circ K=K^{\prime }\circ f  \label{1.6}
\end{equation}

As the composition of two morphisms is a morphism and the identity $I_{M}$
is a morphism, we get a {\bf category of reversal systems}, whose objects
are the reversal systems and whose morphisms are the morphisms of reversal
systems. Also we have the subcategories of symplectic, almost complex,
K\"{a}hlerian, etc. reversal systems.

\section{Abstract reversals and abstract time-reversal systems}

{\bf Definition:} Let $(M,\mu )$, be a measure space, and let $%
K:M\rightarrow M$ be an isomorphism (mod 0) \cite{2}. We will say that $K$
is an {\bf abstract} {\bf reversal} on $(M,\mu )$ and that $(M,\mu ,K)$ is
an {\bf abstract} {\bf reversal system} if the following conditions are
satisfied:

(a.r.1) $K$ is an involution, i.e. $K^2=I_M$ (mod 0)

(a.r.2) The set $N$ of all fixed points of $K$ is a measurable subset of
null measure of $M,$ and so $\mu [M-N]=\mu [M]$ (In particular, this implies
that $M$ is a non trivial involution, i.e. $K\neq I_{M},$ the identity
function on $M$)

We will say that $N$ is the {\bf invariant} {\bf subset} of the abstract
reversal system.

{\bf Definition:} Let $(M,\mu ,K)$ be an abstract reversal system such that
there is a class ${\cal F}$ of measure preserving flows $\left( S_{t}\right)
_{t\in {\Bbb R}}$ or / and cascades $\left( S_{t}=S^{t}\right) _{t\in {\Bbb Z%
}}$ on $M$ such that, for all $m\in M,$ and all $t$ in ${\Bbb R}$ (or in $%
{\Bbb Z}$) it satisfies (\ref{1.5}). Then, we will say that $K$ is a {\bf %
time reversal }for ${\cal F}${\bf \ }on $(M,\mu )$.

Only having an abstract time-reversal on $M,$ {\bf time-symmetric} {\bf (or
asymmetric)} abstract dynamical systems $\left( S_{t}\right) $ (flows $%
\left( S_{t}\right) _{t\in {\Bbb R}}$ ; or cascades $\left(
S_{t}=S^{t}\right) _{t\in {\Bbb Z}}$ ) can be defined. In fact, $\left(
S_{t}\right) $ will be regarded as{\bf \ symmetric with respect to the
time-reversal }$K${\bf ,} if it fulfills $\forall m\in M$ the condition (\ref
{1.5}) (or {\bf asymmetric }if it doesn't)

{\bf Definition: }By a {\bf morphism} of the abstract reversal system $%
(M,\mu ,K)$ into $(M^{\prime },\mu ^{\prime },K^{\prime }),$ we mean a
measurable map $f$ of $(M,\mu )$ into $(M^{\prime },\mu ^{\prime })$ such
that, $\forall A^{\prime }\subset M^{\prime }$ measurable: 
\begin{equation}
\mu \left( f^{-1}(A^{\prime })\right) =\mu ^{\prime }\left( A^{\prime
}\right) \text{ mod 0}  \label{1.7}
\end{equation}
and 
\begin{equation}
f\circ K=K^{\prime }\circ f  \label{1.8}
\end{equation}

As the composition of two morphisms is a morphism and the identity $I_{M}$
is a morphism, we get a {\bf category of abstract reversal systems}, whose
objects are the abstract reversal systems and whose morphisms are the
morphisms of abstract reversal systems.

\section{Examples of time-reversal systems}

We will see how the mathematical structure just described can be implemented
in all the classical and quantum physical systems, and also generalized to
more abstract purely mathematical dynamical systems, as the Bernouilli
systems.

\subsection{The real line}

Let us consider in the real line ${\Bbb R}$, the mapping $K:{\Bbb R}%
\rightarrow {\Bbb R}$ defined by: 
\begin{equation}
K(t)=-t  \label{2.1}
\end{equation}

Clearly, ${\Bbb R}$ is $K$-orientable, because if $N=\{0\},$ then ${\Bbb R}%
-\{0\}={\Bbb R}_{+}\cup {\Bbb R}_{-},$ and $K$ is a canonical time-reversal
for the family of translations: for $a\in {\Bbb R}$ fixed, and $t\in {\Bbb R}
$, 
\begin{equation}
\text{ }S_{t}^{a}(x):=x+ta  \label{2.2}
\end{equation}

\subsection{The Minkowski space-time}

Let $({\Bbb R}^4,\eta )$ be the Minkowski space-time, with $\eta =$ diag $%
(1,-1,-1,-1)$. The invariant submanifold is the spacelike hyperplane 
\[
N=\left\{ (0,x,y,z):x,y,z\in {\Bbb R}\right\} 
\]

Clearly, fixing $M_{+}$ as the halfspace containing the ''forward'' light
cone 
\[
\left\{ (ct,x,y,z):c^{2}t^{2}-x^{2}-y^{2}-z^{2}>0\text{ and }t>0\right\} 
\]
and $M_{-}$ as the halfspace containing the ''backward'' light cone 
\[
\left\{ (ct,x,y,z):c^{2}t^{2}-x^{2}-y^{2}-z^{2}>0\text{ and }t<0\right\} 
\]
and defining $K:{\Bbb R}^{4}\rightarrow {\Bbb R}^{4}$ by: 
\begin{equation}
K(ct,x,y,z)=(-ct,x,y,z)  \label{2.3}
\end{equation}
we get a canonical $K$-orientation, equivalent to the ussual
time-orientation. $K$ is a time-reversal with respect to the temporal
translations 
\begin{equation}
S_{t}^{A}(X):=X+tA=(x^{0}+ta^{0},x^{1},x^{2},x^{3})  \label{2.4}
\end{equation}
for $A=(a^{0},0,0,0)\in {\Bbb R}^{4}:a^{0}\neq 0$ fixed, and $t\in {\Bbb R}$.

{\bf Remark: }As an effect of curvature, not every general 4-dimensional
Lorentzian manifold $(M,g),$ will be time-orientable \cite{Licner}.
Nevertheless, a time-orientable general space-time is necessary if we are
searching for a model of a universe with an arrow of time \cite{cosmic arrow}
\cite{Casta}. In fact, if our universe were represented by a
non-time-orientable manifold, it would be impossible to define past and
future in a global sense, in contradiction with all our present cosmological
observations. Precisely, we know that there are no parts of the universe
where the local arrow of time points differently from our own arrow.

\subsection{The cotangent fiber bundle. Classical Mechanics.}

A general cotangent fiber bundle needs not to be $K$-orientable.
Nevertheless, we have the following result:

{\bf Theorem}: The cotangent fiber bundle (of a finite dimensional
differentiable manifold) $\left( T^{*}(N),\pi ,N\right) $ \cite{1} has a
canonical almost K\"{a}hlerian time-reversal (for the Hamiltonians flows on
it).

{\bf Proof:} Let $M$ be the cotangent fiber bundle $T^{*}(N)$ of a real
n-dimensional manifold $N.$ In this case $N$ is an embedded submanifold of $%
M $, being the embedding $i:N\rightarrow T^{*}(N)$ such that: 
\[
\text{ }i(q)=0_{q}\text{ (the null functional at }q\text{)} 
\]

Let's define

\begin{eqnarray}
K &:&T^{*}(N)\rightarrow T^{*}(N)  \nonumber  \label{4.1} \\
\forall p_q &\in &T_q^{*}(N):K(p_q)=-p_q  \label{2.5}
\end{eqnarray}

Because of its definition, this map is obviously an involution, and by its
linearity is differentiable and its differential or tangent map 
\[
K_{*}:T\left( T^{*}(N)\right) \rightarrow T\left( T^{*}(N)\right) 
\]
verifies: 
\begin{equation}
K_{*}(X_{p_{q}})= 
{X_{p_{q}}\text{ if }X_{p_{q}}\in i_{*}\left( T_{q}(N)\right)  \atopwithdelims\{. -X_{p_{q}}\text{ if }X_{p_{q}}\in T_{p_{q}}\left( \pi ^{-1}(q)\right) }%
\label{2.6}
\end{equation}
$X_{p_{q}}\in T_{p_{q}}\left( \pi ^{-1}(q)\right) $ means that it is
``vertical'' or tangent to the point $p_{q}$ of the fibre in $q$. It must be
taken into account that by joining a vertical base with the image of a base
in $N$ by the isomorphism $i_{*}$ $,$ we get a base of $T_{p_{q}}\left(
T^{*}(N)\right) .$

Let $\omega $ be the canonical symplectic 2-form of the cotangent fiber
bundle. As $\omega $ is antisymmetric, in order to evaluate $K^{*}\omega ,$
it is sufficient to consider only three possibilities: 
\begin{eqnarray*}
1)\;\left( X_{p_q},Y_{p_q}\right) &:&X_{p_q}\;,Y_{p_q}\in i_{*}\left(
T_q(N)\right) \\
2)\;\left( X_{p_q},Y_{p_q}\right) &:&X_{p_q}\;,Y_{p_q}\in T_{p_q}\left( \pi
^{-1}(q)\right) \\
3)\;\left( X_{p_q},Y_{p_q}\right) &:&X_{p_q}\in i_{*}\left( T_q(N)\right) 
\text{ but }Y_{p_q}\in T_{p_q}\left( \pi ^{-1}(q)\right)
\end{eqnarray*}

In case 1) 
\begin{equation}
\omega \left( K_{*}(X_{p_q}),K_{*}(Y_{p_q})\right) =\omega \left(
X_{p_q},Y_{p_q}\right) =0  \label{2.7}
\end{equation}

In case 2) 
\begin{equation}
\omega \left( K_{*}(X_{p_q}),K_{*}(Y_{p_q})\right) =\omega \left(
-X_{p_q},-Y_{p_q}\right) =\omega \left( X_{p_q},Y_{p_q}\right) =0
\label{2.8}
\end{equation}

In case 3) 
\begin{eqnarray}
\omega \left( K_{*}(X_{p_q}),K_{*}(Y_{p_q})\right) &=&\omega \left(
X_{p_q},-Y_{p_q}\right)  \nonumber \\
&=&-\omega \left( X_{p_q},Y_{p_q}\right)  \label{2.9}
\end{eqnarray}

Thus, in any case 
\begin{equation}
\left( K^{*}\omega \right) \left( X_{p_q},Y_{p_q}\right) =\omega \left(
K_{*}(X_{p_q}),K_{*}(Y_{p_q})\right) =-\omega \left( X_{p_q},Y_{p_q}\right)
\label{2.10}
\end{equation}

\noindent which proves that $K^{*}\omega =-\omega ,$ the (s.r.) property$.$

Now, let us define 
\begin{eqnarray}
J &:&T\left( T^{*}(N)\right) \rightarrow T\left( T^{*}(N)\right)  \nonumber
\\
J(X_{p_q}) &=&Y_{p_q}\Leftrightarrow \omega \left( X_{p_q},Y_{p_q}\right) =1
\label{2.10.1}
\end{eqnarray}
that is to say, $J(X_{p_q})$ is the canonical conjugate of $X_{p_q}.$

Then, by the antisymmetry of $\omega ,$ clearly $J^{2}=-I.$ In addition 
\begin{eqnarray}
\left( K_{*}\circ J\right) (X_{p_{q}}) &=&K_{*}\left( J(X_{p_{q}})\right)
=-J\left( X_{p_{q}}\right)  \nonumber \\
&=&-J\left( K_{*}\left( X_{p_{q}}\right) \right)  \nonumber \\
&=&\left( -J\circ K_{*}\right) (X_{p_{q}})  \label{2.10.2}
\end{eqnarray}
if $X_{p_{q}}\in i_{*}\left( T_{q}(N)\right) $ and 
\begin{eqnarray}
\left( K_{*}\circ J\right) (X_{p_{q}}) &=&K_{*}\left( J(X_{p_{q}})\right)
=J(X_{p_{q}})  \nonumber \\
&=&J\left( -K_{*}\left( X_{p_{q}}\right) \right)  \nonumber \\
&=&\left( -J\circ K_{*}\right) (X_{p_{q}})  \label{2.10.3}
\end{eqnarray}
if $X_{p_{q}}\in T_{p_{q}}\left( \pi ^{-1}(q)\right) .$ So, $K$ preserves
both $\omega $ and $J$, and therefore is almost K\"{a}hlerian. $\Box $

As it is well known, the phase space $M$ of a classical system with a finite
number ($n$) of degrees of freedom and holonomic constraints has the
particular form $T^{*}(N),$ where $N$ denotes the configuration space of the
system. This implies the existence of a privileged submanifold $N$ of $M$.
We may enquire: why is this so? The answer is: because every law of
Classical Mechanics is invariant with respect to the Galilei group {\it %
which contains all the spatial translations} (and it is itself a contraction
of the inhomogeneous Lorentz group \cite{Hermann}). This forces the
configuration space to be a submanifold of some {\it homogeneuos} ${\Bbb R}%
^{d}$ space. Now, in this submanifold we also have a privileged system of
coordinates: the spatial position coordinates $q_{1}=x_{1},...,q_{d}=x_{d}$,
with respect to which the action of the Galilei group has its simplest
affine expression. Nevertheless, in general this action will take us away
from the configuration manifold $N,$ because it doesn't fit with the
constraints (think for example in the configuration space of a double
pendulum -with two united threads- which is a 2-torus, contained in ${\Bbb R}%
^{3}$).

\subsection{Classical Statistical Mechanics}

Let us consider the phase space of a classical system $\left(
T^{*}(N),\omega \right) $ and take the real Banach space $V=L_{{\Bbb R}%
}^1\left( T^{*}(N),\sigma \right) $ containing the probability densities
over the phase space, where $\sigma =\omega \wedge ...\wedge \omega $ ($n$
times) is the Liouville measure. $V$ is a real infinite dimensional
differentiable manifold modelled by itself. Then, the above defined almost
K\"{a}hlerian time-reversal $K$ on $T^{*}(N)$ induces $\widetilde{K}%
:V\rightarrow V$ by: 
\begin{equation}
\rho \mapsto \widetilde{K}(\rho ):\left( \widetilde{K}(\rho )\right)
(m):=\rho \left( K(m)\right)  \label{2.11}
\end{equation}

Clearly, $\widetilde{K}$ is a toplinear involution. Now, let us consider the
set $P$ of all (``almost everywhere'' equivalent classes of) ``$\widetilde{K}
$-even'' integrable functions 
\begin{equation}
P=\left\{ \rho \in V:\rho \left( K(m)\right) =\rho (m)\right\}  \label{2.12}
\end{equation}
and the set $I$ of all (``almost everywhere'' equivalent classes of) the ``$%
\widetilde{K}$-odd'' integrable functions 
\begin{equation}
I=\left\{ \rho \in V:\rho \left( K(m)\right) =-\rho (m)\right\}  \label{2.13}
\end{equation}

Trivially, $V=P\oplus I$ , and there are two toplinear projectors mapping
any $\rho \in V$ into its ``$\widetilde{K}$-even'' and ``$\widetilde{K}$%
-odd'' parts. $P$ is the invariant subspace of $\widetilde{K}.$ Its
complement is the (infinite dimensional) open submanifold of (``almost
everywhere'' equivalent classes of) integrable functions whose ``$\widetilde{%
K}$-odd'' projection doesn't vanish$.$

Now, every dynamical system $(S_t)$ (in particular those of the class ${\cal %
F}$ of $K$) on $T^{*}(N)$ induces another dynamical system $(U_t)$ on $V:$%
\begin{equation}
\left( U_t(\rho )\right) (m):=\rho \left( S_{-t}(m)\right)  \label{2.14}
\end{equation}

Considering the class $\widetilde{{\cal F}},$ induced by ${\cal F},$ we
conclude that $\widetilde{K}$ is a time-reversal.

So we have another example lacking time orientability but having a
time-reversal structure .

\subsection{Complex Banach spaces}

A {\bf complex structure} on a {\it real} finite (or infinite) Banach space $%
V$ is a linear (toplinear) transformation $J$ of $V$ such that $J^2=-I $ ,
where $I$ stands for the identity transformation of $V$. \cite{4}

In the case of a {\it complex} Banach space $V_{{\Bbb C}}$ , we can consider
the associated real vector space (its ``realification'') $V=V_{{\Bbb R}}$
composed of the same set of vectors, but with ${\Bbb R}$ instead of ${\Bbb C}
$, as the field of its scalars. Then, $J=iI$ is the canonical complex
structure of $V_{{\Bbb R}}.$

If $J$ is a complex structure on a finite dimensional real vector space, its
dimension must be even. In any case, there exist elements $X_1$, $X_2$,..., $%
X_n,...$ of $V$ such that 
\[
\left\{ X_1,...,X_n,...,JX_1,...JX_n,...\right\} 
\]
is a basis for $V$ \cite{4}.

Let us define $K:V\rightarrow V$ as the ``conjugation'', i.e. extending by
linearity the assignment 
\begin{equation}
\forall i=1,2,...:K(X_{i})=X_{i}\;;\;K(JX_{i})=-JX_{i}  \label{2.15}
\end{equation}

Then the real subspace generated by $\{X_{1},X_{2},...\},$ is the subspace
of fixed points of $K,$ $N$. So, $(V,J,K)$ is a complex time-reversal system
for the class of ''non real translations'' $\left( S_{t}^{A}\right) _{t\in 
{\Bbb R}}$ ($A$ being a linear combination of $JX_{1},...JX_{n},...)$: 
\begin{equation}
S_{t}^{A}(X)=X+tA\;,\;t\in {\Bbb R}  \label{2.16}
\end{equation}

In fact, 
\begin{eqnarray}
\left( K\circ S_{t}^{A}\circ K\right) (X) &=&K\left( K(X)+tA\right) =X-tA 
\nonumber \\
&=&S_{-t}^{A}(X)  \label{2.17}
\end{eqnarray}

\subsection{Ordinary quantum mechanical systems}

As a particular case of the previous example, let us consider a classical
system whose phase space is ${\Bbb R}^{6n},$ and take $V={\cal H}%
=L^{2}\left( {\Bbb R}^{3n},\sigma \right) $ (actually, its realification).
This choice is motivated by the fact that we want to have a
Galilei-invariant Quantum Mechanics, and so {\it we must quantify the
spatial position coordinates}. There is no cannonical or symplectic symmetry
here. Only acting on the wave functions of the position coordinates the
Wigner time-reversal operator will be expressed as the complex conjugation.
So, we get a complex time-reversal structure.

Now, following \cite{Cire1}, let us consider the real but infinite
dimensional K\"{a}hlerian manifold $\left( {\bf P}({\cal H}),\widetilde{J},%
\widetilde{\omega },g\right) $ of the associated projective space ${\bf P}(%
{\cal H})$ of the Hilbert states space ${\cal H}$ of an ordinary quantum
mechanical system. $J$ is the complex structure of ${\cal H}$, and it is the
local expresion of 
\[
\widetilde{J}:T\left( {\bf P}({\cal H})\right) \rightarrow T\left( {\bf P}(%
{\cal H})\right) 
\]

$\left( {\bf P}({\cal H}),\widetilde{J},\widetilde{\omega },g\right) $ has a
canonical K\"{a}hlerian time-reversal structure. In fact, we define $K:{\cal %
H}\rightarrow {\cal H}$ as in the previous example, and take: 
\[
\widetilde{K}:{\bf P}({\cal H})\rightarrow {\bf P}({\cal H})\text{ by: }%
\widetilde{K}\left[ \psi \right] =\left[ K(\psi )\right] 
\]
then, all the desired properties follows easily.

\subsection{Quantum Statistical Mechanics}

Let $V=L^{1}({\cal H})$ denote the complex Banach space generated by all
nuclear operators on ${\cal H}$ with the trace norm. This set contains the
density operators of Quantum Statistical Mechanics. $V$ is a real infinite
dimensional differentiable manifold modelled by itself. Then, the above
defined complex time-reversal $K$ on ${\cal H}$ induces $\widehat{K}%
:V\rightarrow V$ by: 
\begin{equation}
\widehat{\rho }\mapsto \widehat{K}(\widehat{\rho }):\left( \widehat{K}(%
\widehat{\rho })\right) (\psi ):=\widehat{\rho }\left( K(\psi )\right)
\label{2.18}
\end{equation}

Clearly, $\widehat{K}$ is a toplinear involution. Now, let us consider the
set $R$ of all ``$\widehat{K}$-real'' densities 
\begin{equation}
R=\left\{ \widehat{\rho }\in V:\widehat{\rho }\left( K(\psi )\right) =%
\widehat{\rho }(\psi )\right\}  \label{2.19}
\end{equation}
and the set $I$ of all the ``$\widehat{K}$-imaginary'' densities 
\begin{equation}
I=\left\{ \widehat{\rho }\in V:\widehat{\rho }\left( K(\psi )\right) =-%
\widehat{\rho }(\psi )\right\}  \label{2.20}
\end{equation}

Trivially, $V=R\oplus I$ , and there are two toplinear projectors mapping
any $\rho \in V$ into its ``$\widehat{K}$-real'' and ``$\widehat{K}$%
-imaginary'' parts. $R$ is the invariant subspace of $\widehat{K}.$ Its
complement is the (infinite dimensional) open submanifold of (``almost
everywhere'' equivalent classes of) integrable functions whose ``$\widehat{K}
$-imaginary'' projection is not null$.$

Now, every dynamical system $(S_{t})$ (in particular those of the class $%
{\cal F}$ of $K$) on ${\cal H}$ induces another dynamical system $(U_{t})$
on $V$ by setting$:$%
\begin{equation}
\left( U_{t}(\widehat{\rho })\right) (\psi ):=\widehat{\rho }\left(
S_{-t}(\psi )\right)  \label{2.21}
\end{equation}

Considering the class $\widehat{{\cal F}},$ induced by ${\cal F},$ we
conclude that $\widehat{K}$ is a complex time-reversal. So we have another
example lacking time orientability but having a time-reversal structure.

In both Classical and Quantum Statistical Mechanics, we have used the same
criterium to choose $N$ and ${\cal H}$ respectively. The densities of the
two theories are related by the Wigner integral $W$, which is an essential
ingredient in the theory of the classical limit \cite{Casta-Laura}. In the
one dimensional case, it is the mapping $\widehat{\rho }\mapsto \rho
=W\left( \widehat{\rho }\right) $ given by: 
\begin{equation}
\rho (q,p)=\frac{1}{\pi }\int\limits_{-\infty }^{+\infty }\widehat{\rho }%
(q-\lambda ,q+\lambda )\,e^{2ip\lambda }\,d\lambda  \label{2.21a}
\end{equation}
where $q$ is the spatial position coordinate \footnote{%
We want to emphasize the necessity of having an homogeneous configuration
space (${\Bbb R}$ in the one dimensional case) in order to have the
translations $q\mapsto q\pm \lambda $ in $W$ integral.}, $p$ its conjugate
momentum, $\rho (q,p)$ a classical density function, and 
\begin{eqnarray}
\widehat{\rho }(x,x^{\prime }) &=&\left( \sum\limits_{j=1}^{\infty }\rho
_{j}\,\overline{\psi _{j}}\otimes \psi _{j}\right) (x,x^{\prime })  \nonumber
\\
&=&\sum\limits_{j=1}^{\infty }\rho _{j}\,\overline{\psi _{j}}(x)\psi
_{j}(x^{\prime })  \label{2.21b}
\end{eqnarray}
is a generic matrix element of a quantum density, being $\left\{ \psi
_{j}\right\} _{j=1}^{\infty }$ an orthonormal base of ${\cal H}$, $\rho
_{j}\geqslant 0$ and $\sum\limits_{j=1}^{\infty }\rho _{j}=1$.

As it is obvious by a simple change of variables, 
\begin{equation}
W\left[ \widehat{K}\left( \widehat{\rho }\right) \right] \left( q,p\right)
=\rho (q,-p)=\rho \left( K(q,p)\right) =\widetilde{K}\left[ W\left( \widehat{%
\rho }\right) \right] \left( q,p\right)  \label{2.21c}
\end{equation}
and therefore, $W$ is a morphism between $\left( L^{1}({\cal H}),\widehat{K}%
\right) $ and $\left( L_{{\Bbb R}}^{1}\left( T^{*}(N)\right) ,\widetilde{K}%
\right) .$

\subsection{Koopman treatment of Kolmogorov-Systems}

With the definition of time-reversal in the physical examples above, we now
face the same definition in purely mathematical dynamical systems.

Let $(M,\mu ,S_t)$ be a Kolmogorov system (cascade or flow). As it is well
known, this implies that the induced unitary evolution $U_t$ in ${\cal H}%
=[1]^{\bot }$ the orthogonal complement of the one dimensional subspace of
the classes a. e. of the constant functions in the Hilbert space $L^2(M,\mu
),$ has uniform Lebesgue spectrum of numerable constant multiplicity. This,
in turn, implies the existence of a system of imprimitivity $(E_s)_{s\in 
{\Bbb G}}$ based on ${\Bbb G}$ for the group $(U_t)_{t\in {\Bbb G}}$ , where 
${\Bbb G}$ is ${\Bbb Z}$ or ${\Bbb R}$: 
\begin{equation}
E_{s+t}=U_tE_sU_t^{-1}  \label{2.22}
\end{equation}

Following Misra \cite{Misra}, we define the ``Aging'' operator 
\begin{equation}
T=\int\limits_{{\Bbb G}}s\,dE_{s}= 
{\int\limits_{{\Bbb R}}s\,dE_{s}\text{ for fluxes} \atopwithdelims\{. \sum\limits_{s\in {\Bbb Z}}sE_{s}\text{ for cascades}}%
\label{2.23}
\end{equation}

Then 
\begin{equation}
U_{-t}TU_t=T+t  \label{2.24}
\end{equation}

$T$ is selfadjoint in the discrete case, and essentially selfadjoint in the
continuous case, and there are eigenvectors in the discrete case, and
generalized eigenvectors (antifunctionals) in certain riggings of ${\cal H}$
by a nuclear space $\Phi $ ($\Phi \prec {\cal H}\prec \Phi ^{\times }$) in
the continuous case, $\left( \left| \tau ,n\right\rangle \right) _{\tau \in 
{\Bbb G}}$ , such that: 
\begin{eqnarray}
T\left| \tau ,n\right\rangle &=&\tau \left| \tau ,n\right\rangle
\label{2.25} \\
U_{t}\left| \tau ,n\right\rangle &=&(\tau +t)\left| \tau ,n\right\rangle
\label{2.26}
\end{eqnarray}

Defining 
\begin{equation}
K\left| \tau ,n\right\rangle =-\left| \tau ,n\right\rangle  \label{2.27}
\end{equation}

It follows easilly that $K$ restricted to ${\cal H}$ is a time-reversal for $%
{\cal F}=\{(U_t)\}$ with respect to which $U_t$ is symmetric.

\section{Examples of abstract time-reversals}

\subsection{Bernouilli schemes}

Let $M$ be the set $\Sigma ^{{\Bbb Z}}$ of all bilateral sequences (of
``bets'') 
\begin{equation}
m=(a_{j})_{j\in {\Bbb Z}}=(...a_{-2},a_{-1},a_{0},a_{1},a_{2},...)
\label{3.1}
\end{equation}
on a finite set $\Sigma $ with $n$ elements (a ``dice'' with $n$ faces). Let 
${\frak X}$ be the $\sigma $-algebra on $M$ generated by all the subset of
the form 
\begin{equation}
A_{j}^{s}=\{m:a_{j}=s\in \Sigma \}  \label{3.2}
\end{equation}

Clearly, 
\begin{equation}
M=\bigcup\limits_{s\in \Sigma }A_j^s=\bigcup\limits_{k=1}^nA_j^{s_k}
\label{3.3}
\end{equation}

Let's define a normalized measure $\mu $ on $M$ by choosing $n$ ordered
positive real numbers $p_{1},...,p_{n}$ whose sum is equal to one ($p_{k}$
is the ``probability'' of getting $s_{k}$ when the ``dice'' is thrown), and
setting: 
\begin{equation}
\forall k:k=1,...,n\;:p_{k}=\mu (A_{j}^{s_{k}})  \label{3.4}
\end{equation}

\begin{equation}
\mu \left( A_{j_1}^{s_1}\cap ...\cap A_{j_k}^{s_k}\right) =\mu
(A_{j_1}^{s_1})...\mu (A_{j_k}^{s_k})  \label{3.5}
\end{equation}
where $j_1,...,j_k$ are all different.

Let the dynamical authomorphism $S$ be the shift to the right: 
\begin{eqnarray}
S\left( (a_{j})_{j\in {\Bbb Z}}\right) &=&(a_{j}^{\prime })_{j\in {\Bbb Z}} 
\nonumber \\
\text{where: } &&a_{j}^{\prime }:=a_{j-1}  \label{3.6}
\end{eqnarray}

The shift preserves $\mu $ because 
\begin{equation}
\mu \left( S(A_j^{s_k})\right) =\mu (A_{j+1}^{s_k})=p_k  \label{3.7}
\end{equation}

The above abstract dynamical scheme is called a Bernouilli scheme and
denoted $B(p_1,...,p_n).$

Let's define a ``cannonical'' abstract reversal by: 
\begin{eqnarray}
K\left( (a_{j})_{j\in {\Bbb Z}}\right) &=&(a_{j}^{\prime })_{j\in {\Bbb Z}} 
\nonumber \\
a_{j}^{\prime } &=&a_{-j+1}  \label{3.8}
\end{eqnarray}

Clearly, $K$ is an isomorphism, and its invariant set 
\begin{equation}
N=\bigcap\limits_{j\in {\Bbb Z}}\left\{ \bigcup\limits_{s\in \Sigma }\left(
A_j^s\cap A_{-j}^s\right) \right\}  \label{3.9}
\end{equation}
has $\mu $-measure $0.$ In addition $K$ is a time-reversal for the class $%
{\cal F}$ of all Bernouilli schemes, because 
\begin{equation}
K\circ S\circ K=S^{-1}  \label{3.10}
\end{equation}
being $S^{-1}$ the shift to the left.

\subsection{The Baker's transformation}

We will show the geometrical meaning of the last two time-reversals for the
Baker's transformation.

The measure space is the torus 
\[
M=\left[ 0,1\right] \times \left[ 0,1\right] \;/\sim \;=\left\{ (x,y)%
\mathop{\rm mod}%
1=[x,y]:x,y\in \left[ 0,1\right] \right\} 
\]
that is to say, $\sim $ is the equivalence relation that identifies the
following boundary points: 
\[
(0,x)\sim (1,x)\;\text{and}\;(x,0)\sim (x,1) 
\]
with its Lebesgue measure. The automorphism $S$ acts as follows: 
\begin{equation}
S(x,y)= 
{(2x,\frac 12y)\;\;\;\;\;\;\;\;\;\;if\;0\leq x\leq \frac 12\;,\;0\leq y\leq 1 \atopwithdelims\{. (2x-1,\frac 12y+\frac 12)\;if\;\frac 12\leq x\leq 1\;,\;0\leq y\leq 1}%
\label{5.1}
\end{equation}

It's clear that $S$ is a non-continuous{\bf \ }but measure preserving
transformation which involves a contraction in the $y$ direction and a
dilatation in the $x$ direction: the contracting and dilating directions at
every point $m\in M$ (that is, the vertical and the horizontal lines through
each $m$).

The torus is a compact, connected Lie group and we can define an involutive
automorphism $K$ on $M$ by putting:

\begin{equation}
K[x,y]=[y,x]\;;\;x,y\in I=\left[ 0,1\right]  \label{5.2}
\end{equation}

The fixed points of $K$ constitute a submanifold of the torus: the
projection of the diagonal $\Delta $ of the unit square $I\times I$

\[
N=\Delta \;/\sim =\left\{ [x,x]:x\in I\right\} 
\]

Then: 
\begin{equation}
K\circ S^t\circ K=S^{-t},\;\forall t\in {\Bbb Z}  \label{5.3}
\end{equation}

In fact, the first application of $K$ to the generating partition of $S$,
rotates the unit square, interchanging the $x$ fibers with the $y$ ones.
Then by applicating $S^{t}$ (that is $t$ times $S$) we get a striped pattern
of horizontal lines, which is rotated and yields a striped pattern of
vertical lines when $K$ is applicated again. The same pattern would be
obtained if $S^{-t}$ was used.

As it is well known \cite{2}, the Baker transformation is isomorphic to $B(%
\frac{1}{2},\frac{1}{2})$. In fact, the map 
\begin{equation}
(x,y)\mapsto m=(a_{j})_{j\in {\Bbb Z}}\Leftrightarrow
x=\sum\limits_{j=0}^{\infty }\frac{a_{-j}}{2^{j+1}}\;\text{and }%
y=\sum\limits_{j=1}^{\infty }\frac{a_{j}}{2^{j}}  \label{5.4}
\end{equation}
is an isomorphism (mod 0). Moreover it is an isomorphism of abstract
time-reversal systems, because it sends the time-reversal of (\ref{5.2}) in
the time-reversal of (\ref{3.8}). In particular, this implies that the
Baker's map is a Kolmogorov system, and therefore having the corresponding
time-reversal for its Koopman treatment. These three reversals are related.

Let $\{A,B\}$ be the partition of the unit square into its left and right
halves. As it is well known this partition is both independent and
generating for the Baker's map. Let's define 
\begin{equation}
\theta _{0}=1-\chi _{A}=%
{\text{\ }1\text{ in }A \atopwithdelims\{. -1\text{ in }B}%
\label{5.5}
\end{equation}
where $\chi _{A}$ is the characteristic function of the set $A,$ as well as 
\begin{equation}
\theta _{n}=U^{n}(\theta _{0})=\theta _{0}\circ S^{-n}= 
{\text{\ }1\text{ in }S^{n}(A) \atopwithdelims\{. -1\text{ in }S^{n}(B)}%
\label{5.6}
\end{equation}
and for any finite set $F=\{n_{1},...,n_{F}\}\subset {\Bbb Z}$, put 
\begin{equation}
\theta _{F}=\theta _{n_{1}}...\theta _{n_{F}}\;\text{(ordinary product of
functions)}  \label{5.7}
\end{equation}

Then, all the eigenvectors of the Aging operator $T$ of $U$ are of the form 
\cite{Prigo}: 
\begin{equation}
T\theta _{F}=n_{m}\theta _{F}  \label{5.8}
\end{equation}
where $n_{m}=\max F.$ Geometrically speaking, $\rho =\theta _{0}$ -that we
can identify with $\{A,B\}$- is an eigenvector of age $0,$ and if $U$ acts $%
n $ times on it we get an eigenvector of age $n$: $\theta _{n}$ -which can
be identified with a set of horizontal fringes-$.$ On the other hand if $%
U^{-1}$ acts $n$ times on it we get an eigenvector of age $-n$: $\theta
_{-n} $ -which can be identified with a set of vertical fringes-$.$ As
expected, the induced action of $K$ sends the ``future'' horizontal
eigenstates of $T$ to the ``past'' vertical ones, and reciprocally.\newpage

\begin{center}
ACKNOWLEDGMENT
\end{center}

The authors wish to express their gratitude to Dr. Sebastiano Sonego for
providing an initial and fruitful discussion on the subject of this paper.
This work was partially supported by grant PIP 4410 of CONICET (Argentine
National Research Council)

\end{document}